\documentclass[journal,10pt]{IEEEtran}

\usepackage{cite}
\usepackage{amsmath,amssymb,amsfonts}
\usepackage{algorithmic}
\usepackage{graphicx}
\usepackage{textcomp}
\usepackage{xcolor}
\usepackage{amsthm}
\theoremstyle{definition}   
\newtheorem{mytheorem}{Theorem}

\begin{document}

\title{Study on MCS Selection and Spectrum Allocation for URLLC Traffic under Delay and Reliability Constraint in 5G Network}

\author{Yuehong Gao, Changhao Sun, Xiaonan Zhang and Xiao Hong
\thanks{Y. Gao, C. Sun, X. Zhang and X. Hong are with the School of Information  and Communication Engineering, Beijing University of Posts and Telecommunications, Beijing 100876, China (email: yhgao@bupt.edu.cn).}
}
\markboth{   }
{}



\maketitle

\begin{abstract}
To support Ultra-Reliable and Low Latency Communications (URLLC) is an essential character of the 5th Generation (5G) communication system. Unlike the other two use cases defined in 5G, e.g. enhanced Mobile Broadband (eMBB) and massive Machine Type Communications (mMTC), URLLC traffic has strict delay and reliability requirement. In this paper, an analysis model for URLLC traffic is proposed  from the generation of a URLLC traffic until its transmission over a wireless channel, where channel quality, coding scheme with finite coding length, modulation scheme and allocated spectrum resource  are taken into consideration. Then, network calculus analysis is applied to derive the delay guarantee for periodical URLLC traffic. Based on the delay analysis, the admission region is found under certain delay and reliability requirement, which gives a lower bound on required spectrum resource. Theoretical results in the scenario of a 5G New Radio system are presented, where the SNR thresholds for adaptive modulation and coding scheme selection, transmission rate and delay, as well as admission region  under different configurations are discussed.  In addition, simulation results are obtained and compared with theoretical results, which validates that the admission region derived in this work provides a lower spectrum allocation bound.
\end{abstract}


\IEEEpeerreviewmaketitle

\section{Introduction}

As it is widely known that 5th Generation (5G) communication network will be able to support three key usage scenarios, i.e., enhanced Mobile Broadband (eMBB), massive machine type communication (mMTC), and Ultra-Reliable and Low-Latency Communication (URLLC)\cite{Osseiran:COMMAG}. 
eMBB aims to support stable connections with high data rates, mMTC is used to support massive connections with low transmission rate and power, while URLLC is proposed to support fast data transmission. Different with eMBB and mMTC, the data payload of URLLC is relatively small but with strict delay and reliability guarantee requirement. As mentioned in \cite{3GPP:38913}, a general URLLC reliability requirement for one transmission of a packet is $1-10^{-5}$ for $32$ bytes with a user plane latency of $1$ms. In addition, not only in 5G system, but also in future 6G system, even more stringent latency and reliability guarantee are need\cite{Kato:6G}. Therefore, how to fulfill such strict quality of service (QoS) requirement for URLLC traffic becomes a quite challenging issue in 5G network.

There have been hot discussions on URLLC, because it is crucial for some important deployment scenarios in 5G, such as vehicular-to-everything (V2X) communication and internet-of-things (IoT) communication.
In \cite{Liu:URLLCV2V}, the transmission of URLLC in a vehicle-to-vehicle network network is considered, where the two queue-aware power allocation solutions are proposed under the constraint of finite block-length.
The authors of \cite{Kato:SmartCity} focus on different QoS requirement of various services in IoT-based smart city, and propose an optimal edge resource allocation to minimize the average service response time.
In addition, multiplexing of URLLC with eMBB traffic is also a key problem to be studied. The authors of \cite{Pocovi:JointScheduling} proposed a joint link adaptation and resource allocation policy, which dynamically adjusts the block error probability of URLLC small payload transmissions in accordance with the instantaneous experienced load per cell with/without eMBB traffic.

There are also many research on resource allocation optimization or key techniques for URLLC traffic.
In \cite{Jacobsen:Grantfree}, the grant-free approach is analyzed and compared with grant-based approach for the uplink transmission of URLLC, where the impact of HARQ mechanisms are also studied.
In \cite{Han:Reservation}, a mechanism of deterministic resource allocation is proposed to meet the reliability and latency requirements, where resource allocation and modulation and coding (MCS) schemes selection are jointly considered.
A novel sparse vector coding (SVC) scheme is proposed in \cite{Kim:TVT} to further improve the reliability of short packet transmission.
The authors of \cite{Xiang:TVT} propose a novel processing architecture capable of performing reception, Orthogonal Frequency Division Multiplexing (OFDM) demodulation and turbo decoding concurrently, rather than consecutively.

Based on these literatures, it is not difficult to notice that every factor during the transmission of a URLLC message will influence the final delay and reliability guarantee, because the QoS requirement is really strict.  Therefore, a thorough model is needed by involving all possible aspects  from the generation of a URLLC message to its receiving over a uncertain wireless channel.

Different with existing work, this paper will firstly establish an theoretical analysis model, which reveals the relationship among  channel quality, finite coding length, modulation scheme, delay guarantee, reliability requirement and information length. The analysis is not correlated with specific scenario or application, and it finally shows the minimum spectrum allocation bound defined as admission region in this work.
Then, configurations standardized in practical 5G New Radio (NR) network are applied to obtain numerical results, where the Signal to Noise Ratio (SNR) thresholds used for adaptive MCS selection and admission region under certain delay and reliability constraint are presented. Lastly, system level simulation is conducted to validate the theoretical analysis.

\section{System Model}\label{sec:systemmodel}

Consider a wireless downlink transmission link started from an information source, who generates information bits. As suggested in \cite{TR38802}, The arrival of URLLC information message is usually assumed to follow periodical model with fixed length. These bits are then coded into codeword and transmitted through a wireless channel.


Let $k$ denote the length of URLLC information bits and $r_c$ denote the coding rate, the length of the coded symbols (denoted by $n$)  will be
\begin{eqnarray}\label{eq:sigmahat}
n=\frac{k}{r_c}.
\end{eqnarray}


It has been proved that, if the generated information bits (with length of $k$ bits) is firstly encoded into a codeword with length of $n$ symbols, and then transmitted through a wireless channel with bandwidth of $W$ at Signla-to-Noise Ratio (SNR) of $P$, the channel transmission rate is upper bounded by Shannon capacity  as
\begin{equation}\label{eq:ShannonEq}
R(\infty,P,W)\leq  W\cdot C(P),
\end{equation}
where
\begin{equation}\label{eq:CP}
C(P)= \log_2(1+P).
\end{equation}

Note that, the upper bound given by Eq.(\ref{eq:ShannonEq}) requires  Gaussian coding and infinite coding length (i.e. $n\rightarrow \infty$). Therefore, when the coding length is finite, it is not possible to realize reliable transmission with arbitrarily small error probability. In \cite{Polyanskiy:Dispersion}\cite{Polyanskiy:Dispersion2}, the maximal rate achievable with error probability $\epsilon$ and finite coding length $n$ is approximated. For wireless channel, its equivalent baseband model is complex Gaussian channel, which is composed of two real Additive White Gaussian Noise (AWGN) channel. Then, based on the Theorem 4 in \cite{Polyanskiy:Dispersion}, it can be obtained that, the transmission rate under given error probability $\epsilon$ and finite coding length $n$ is upper bounded by
\begin{equation}\label{eq:RateCompAWGN}
R_1(n,P,W,\epsilon) \leq W \cdot R_{ch}^1(n,P,\epsilon),
\end{equation}
where
\begin{equation}
\begin{aligned}
R_{ch}^1(n,P,\epsilon) &= C(P) - \sqrt{\frac{V(P)}{n}}Q^{-1}(\epsilon)+\frac{\log_2(n)}{n}, \\
Q(x)&=\int_x^{\infty}\frac{1}{\sqrt{2\pi}}e^{-t^2/2}dt, \\
V(P)&=\frac{P(P+2)}{(P+1)^2\ln^2 2}.
\end{aligned}
\end{equation}

One important assumption in Eq.(\ref{eq:RateCompAWGN}) is Gaussian coding. However, in practical systems, constellation diagram with limited points is usually used. When considering  the widely used M-QAM modulation, the maximal rate under infinite coding length is no longer $C(P)$ defined in Eq.(\ref{eq:CP}), the following upper bound should be used instead
\begin{equation}\label{eq:RateIP}
R(\infty,P,W,M)\leq  W\cdot I(P,M),
\end{equation}
where
\begin{equation}\label{eq:IP}
\begin{aligned}
I(P,M)&=\log_2 M-\frac{1}{M\pi}\sum_{i=1}^{M}\int e^{-\| y-\sqrt{P}x_i \|^2} \\
& \times \left( \sum_{k=1}^{M} e^{-| y-\sqrt{P}x_i |^2 -| y-\sqrt{P}x_k |^2 }  \right) \text{d}y.
\end{aligned}
\end{equation}

However, there is no closed form expression for Eq. (\ref{eq:IP}). Fortunately, the authors of \cite{Ouyang}\cite{Yangpei} found an approximation based on multi-exponential delay curve fitting (M-EDCF) as
\begin{equation}\label{eq:newIP}
I'(P,M) \approx \log_2 M \times \left( 1- \sum_{j=1}^{k_M} a_j^{(M)} e^{-b_j^{(M)} P}  \right),
\end{equation}
where the fitting coefficients for M-QAM are summarized in Table. \ref{tbl:coefficients}.

Then, for finite coding length and finite constellation, the maximal achievable transmission rate can be approximated as
\begin{equation}\label{eq:RateMQAM}
R_2(n,P,W,\epsilon,M) = W\cdot R_{ch}^2(n,P,\epsilon,M),
\end{equation}
where
\begin{equation}
\begin{aligned}
&R_{ch}^2(n,P,\epsilon,M) \\
&= I'(P,M)- \sqrt{\frac{V(P)}{n}}Q^{-1}(\epsilon)+\frac{\log_2(n)}{n} .
\end{aligned}
\end{equation}

\begin{table*}
  \centering
  \caption{Fitting Coefficients for M-QAM}\label{tbl:coefficients}
\begin{tabular}{c c c c c c c c c c}
  \hline
  \hline
  \centering
  $M$  &  $k_M$  &  $a_1^{(M)}$  & $a_2^{(M)}$  & $a_3^{(M)}$  &  $a_4^{(M)}$  & $b_1^{(M)}$ & $b_2^{(M)}$  &  $b_3^{(M)}$  & $b_4^{(M)}$  \\
  256  &  4  & 0.228768  &  0.229083  & 0.118223   & 0.423927    & 0.183242  & 0.038011   & 0.994472   &  0.006911 \\
  64  &  4  & 0.198324  & 0.512831   & 0.209086   & 0.079759    & 0.408618  & 0.027517   & 0.120616   &  1.467118 \\
  16  &  3  & 0.658747  & 0.117219   & 0.224034   &  --  & 0.115521  & 1.467927   & 0.482023   &  -- \\
  4  &  2  & 0.143281  & 0.856719   & --   &  --   & 1.557531  &  0.57239  & --   & --  \\
  \hline
  \hline
\end{tabular}
\end{table*}

%

\section{Analysis on Delay and Admission Region}

In order to analyze the delay and reliability guarantee of the aforementioned system, a suitable analysis model is needed. In this work, network calculus will be relied on to establish a corresponding mathematical model for the considered system.

\begin{figure}
  \centering
  \includegraphics[width=0.45\textwidth]{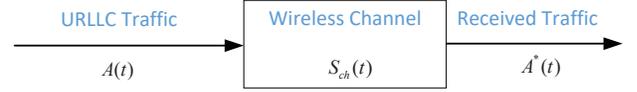}\\
  \caption{Equivalent Analysis Model for the Considered Wireless Communication System}\label{fig:ncmodel}
\end{figure}

Network calculus is a queueing theory for QoS analysis of computer networks\cite{Jiang:StoNetCal}, which has been widely used for performance evaluation in various networks. For the consider URLLC traffic model and wireless channel models described in Section \ref{sec:systemmodel}, an equivalent analysis model can be abstracted as Fig. \ref{fig:ncmodel} shows.  The arrival process, denoted by $A(t)$, is composed of all information bits generated from URLLC information bits; while the service process, denoted by $S_{ch}(t)$, is the service provided by the wireless channel. In network calculus analysis, arrival curve and service curve are defined to describe the characteristics of (cumulative) arrival process and (cumulative) service process, respectively, as follows.

\textbf{Definition 1: Arrival Curve.} An arrival process $A(t)$ is said to have an arrival curve $\alpha(t)$ if for all $0 \leq s \leq t$ \cite{Jiang:StoNetCal}\cite{Jiang:SAC},
\begin{eqnarray}
A(t)-A(s) \leq \alpha(t-s),
\end{eqnarray}
or equivalently $A(t-s) \leq \alpha(t-s)$.

\textbf{Definition 2: Service Curve.} Consider a system $S$ with input process $A(t)$ and output process $A^*(t)$. The system is said to have a service curve $\beta(t)$ if for all $t \geq 0$\cite{Jiang:StoNetCal},
\begin{eqnarray}
A^*(t) \geq A \otimes \beta(t).
\end{eqnarray}

Based on the arrival curve and service curve, the delay guarantee of the received process $A^*(t)$ can be derived by Theorem.\ref{theo:delaybound}.
\begin{mytheorem}\label{theo:delaybound}
If the arrival process has an  arrival curve $\alpha(t)$, and the system provides the arrival process with a service curve $\beta(t)$, then the delay $D(t)$ of the arrival process at time $t$ is bounded by
\begin{eqnarray}
D(t)\leq h(\alpha,\beta ) .
\end{eqnarray}
where $h(\alpha,\beta)$ is the maximum horizontal distance between functions of $\alpha(t)$ and $\beta(t)$, defined as $h(\alpha,\beta)=\sup_{s\geq 0}\left\{ \inf\{\tau\geq 0: \alpha(s)\leq \beta(s+\tau)\}  \right\}$.
\end{mytheorem}

For the considered  URLLC traffic, it is assumed to be a periodical process with interval of $\tau$ and the length of its information bits of $k$, then the arrival curve $\alpha(t)$ can be expressed as
\begin{equation}
\alpha(t)=\frac{k}{\tau}t+k.
\end{equation}

For the considered wireless channel with finite coding length and finite constellation, when bandwidth $W$, SNR $P$, coding rate $r_c$, error probability $\epsilon$ and modulation order $M$ are given, the service process  $S_{ch}(t)$  has service curve $\beta_{ch}(t)$  as:
\begin{equation}
\beta_{ch}(t)= R_{2} \cdot t.
\end{equation}

Then, delay of the received process $A^*(t)$ at the receiver side can be derived as
\begin{equation}\label{eq:delayforperi}
\begin{aligned}
D(t) \leq  &\frac{k}{R_{2}},  \\
\text{Subject to:  } &\frac{k}{\tau} \leq R_{2}.
\end{aligned}
\end{equation}
Detailed proof can be found in \cite{Jiang:StoNetCal}, which is not included in this work for conciseness.

It could be decomposed that the delay guarantee is mainly determined by three parts: 1) how an information message is processed, mainly including coding and modulation schemes; 2) how the channel condition is when it is transmitted, indicated by SNR and error probability; 3) how much system resource is allocated  denoted by system bandwidth.

Usually, delay guarantee and error probability are required by specific URLLC service, the modulation and coding schemes are selected based on channel quality, which are all objective. Therefore, the system can only try it best to allocate reasonable amount of resource in order to provide strict delay and reliability guarantee for URLLC traffic. Given delay and reliability requirement (denoted by $d_0, \epsilon_0$), the minimum required bandwidth is defined as the admission region. Based on the expressions given in \ref{eq:delayforperi}, the delay requirement can be guaranteed as far as allocated system bandwidth is no less than:
\begin{equation}
W_2= \frac{k}{d_0\cdot R_{ch}^2(P,n,\epsilon_0,M)}.
 \end{equation}

\section{Evaluation Results}\label{sec:results}

In 5G New Radio (NR) system, flexible configurations can be applied in order to fulfill different QoS requirements. 3GPP defines $32$  MCS schemes to be used in Physical Downlink Shared CHannel (PDSCH) for 5G NR Rel.15 in Table 5.1.3.1-2 in  \cite{TS38214}. In this work, the following $5$ MCSs with significant different spectrum efficiency will be used in the evaluation as listed in Table. \ref{tbl:mcs}. Note that, the coding rate defined in Table 5.1.3.1-2 in  \cite{TS38214} is actually the binary code rate. For ease of understanding, the coding rate $r_c$ used in Eq.(\ref{eq:sigmahat}) (named as overall code rate in Table. \ref{tbl:mcs}) is also listed out.

\begin{table}
  \centering
  \caption{MCS Configurations}\label{tbl:mcs}
\begin{tabular}{c| c| c | c}
  \hline
  \hline
  \centering
  MCS Index  &  $M$ & Binary Code Rate & Overall Code Rate $r_c$  \\
  \hline
  0     &    4    &    0.11719  &  0.2344                     \\
  5     &    16   &    0.36914  &  1.4766                      \\
  11    &    64   &    0.45508  &  2.7305                    \\
  20    &    256  &    0.66650  &  5.3320                    \\
  27    &    256  &    0.92578  &  7.4063                    \\
  \hline
  \hline
\end{tabular}
\end{table}

\begin{figure}
  \centering
  \includegraphics[width=0.45\textwidth]{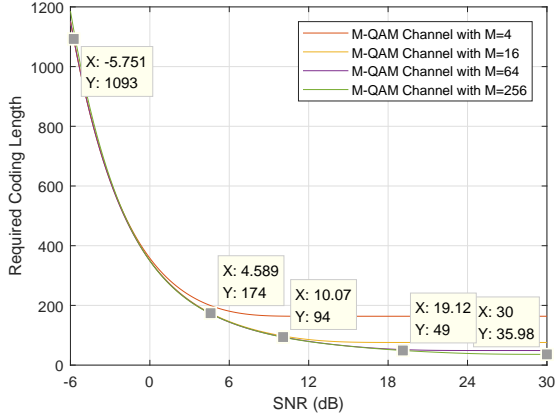}\\
  \caption{Required coding length with $k=256bits$ and $\epsilon=10^{-5}$}\label{fig:codinglength}
\end{figure}

\begin{figure}
  \centering
  \includegraphics[width=0.45\textwidth]{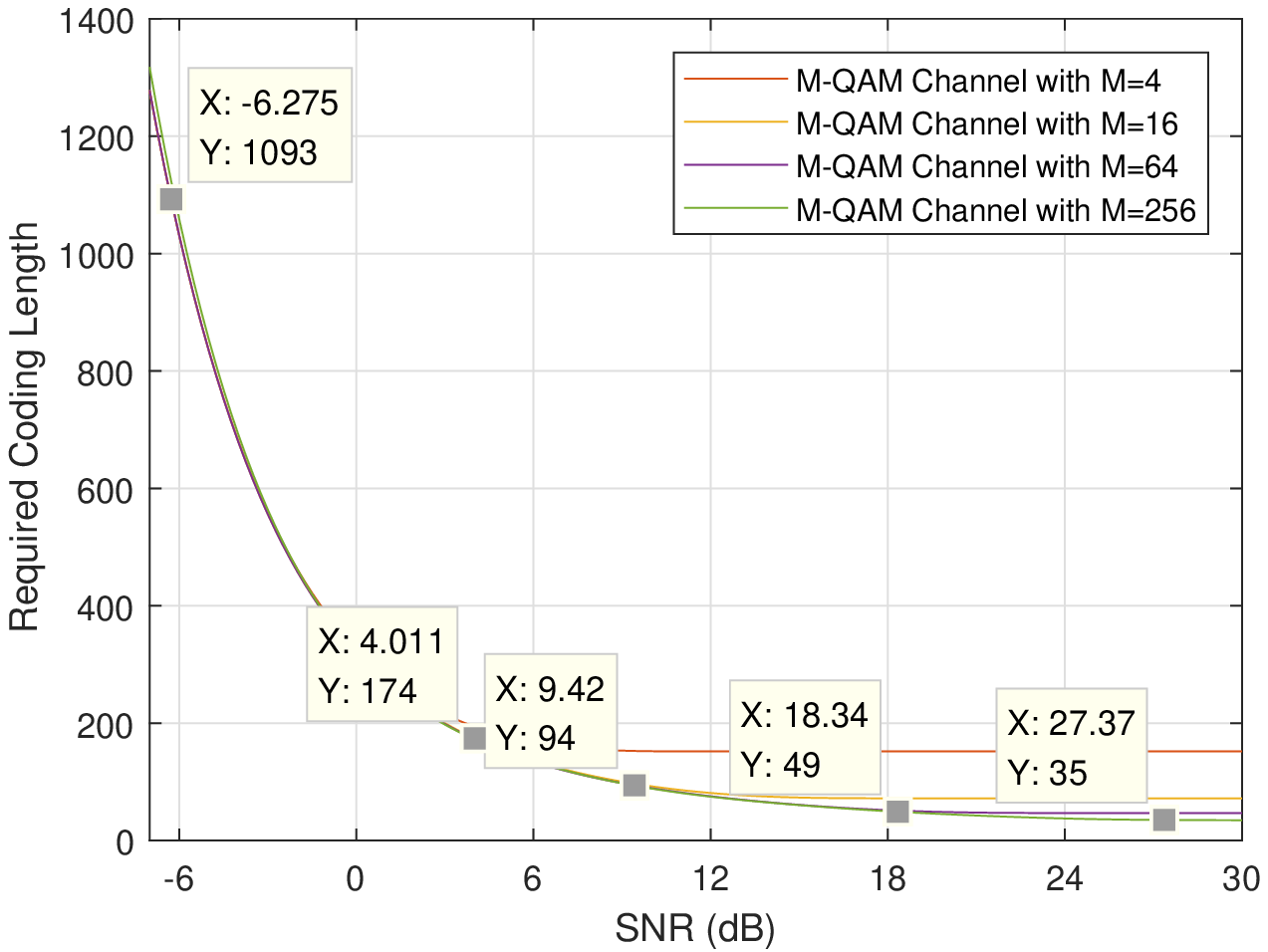}\\
  \caption{Required coding length with $k=256bits$ and $\epsilon=10^{-3}$}\label{fig:codinglength2}
\end{figure}

Firstly, we discuss the minimum required coding length under different SNRs.   Fig. \ref{fig:codinglength} and  Fig. \ref{fig:codinglength2} show the results when the length of URLLC information bits $k$ and error probability $\epsilon$ are set to  $(256, 10^{-5})$  and $(256, 10^{-3})$, respectively.  It is easy to notice that, when channel quality is good (i.e. SNR $P$ is large) and modulation order is high (i.e. $M$ is large), the required coding length will be short. These two figures plot the theoretical lower bound on coding length. When considering the practical MCS schemes defined in Table. \ref{tbl:mcs}, the application feasibility and condition should be carefully discussed.

It is straightforward to calculate the practical coding lengths, denoted by $\hat{n}$ under different MCS indexes by applying Eq.(\ref{eq:sigmahat}) when information length is given. Table. \ref{tbl:codinglength} lists the coding length  when $k=256$bits.

\begin{table}
  \centering
  \caption{Practical Coding Length in 5G NR (when $k=256$ bits)}\label{tbl:codinglength}
\begin{tabular}{c| c}
  \hline
  \hline
  \centering
  MCS Index  &  Coding Length $\hat{n}$  \\
  \hline
  0     &    1093    \\
  5     &    174    \\
  11    &    94   \\
  20    &    48   \\
  27    &    35   \\
  \hline
  \hline
\end{tabular}
\end{table}

If we mark the practical coding length on the corresponding curve given in Fig. \ref{fig:codinglength} and  Fig. \ref{fig:codinglength2}, the minimum required SNR can be found. To be specific, when MCS index $0$ is applied, SNR should be no lower than $-5.751$dB in order to guarantee error probability of $10^{-5}$, and SNR no lower than $-6.275$dB in order to guarantee error probability of $10^{-3}$. Another important issue to be noted here is the point $(X=30,Y=35.98)$, which means the required coding length is $35.98$ when SNR is $30dB$. However, the practical coding length is only $35$. In other words, error probability  of $10^{-5}$ can not be guaranteed when applying MCS Index $27$ even with high SNR of $30$dB (or even upto $50$dB which is not shown in this figure). For comparison, if error probability is lowered to $10^{-3}$, MCS Index $27$ can be applied when SNR is higher than $27.37$dB as marked in Fig.\ref{fig:codinglength2}.

Based on the marks given in Fig. \ref{fig:codinglength} and  Fig. \ref{fig:codinglength2}, SNR thresholds for adaptive MCS selection can be found as listed in Table. \ref{tbl:SNRthresholds}. When the channel quality varies, suitable MCS scheme can be selected dynamically in order to achieve better performance.

\begin{table*}[!t]
  \centering
  \caption{SNR Thresholds for MCS Selection (when $k=256$ bits)}\label{tbl:SNRthresholds}
\begin{tabular}{c| c| c| c| c| c}
  \hline
  \hline
  \centering
  MCS Index  &  1   &   5   &   11  & 20  &   27  \\
  \hline
  SNR Range (dB) ($\epsilon=10^{-5}$)  & [-5.751, 4.589)  & [4.589, 10.07) & [10.07, 19.12) & [19.12, $\infty$)  & $\diagup$   \\
  SNR Range (dB) ($\epsilon=10^{-3}$)  & [-6.275, 4.011)  & [4.011, 9.42) & [9.42, 18.34) & [18.34, 27.37)  & [27.37, $\infty$)  \\
  \hline
  \hline
\end{tabular}
\end{table*}

Given  allocated bandwidth, transmission rates under different MCS schemes can be obtained. In the frequency domain of NR system, a Resource Block (RB) is typically composed of $12$ sub-carriers with interval of $15$kHz, which leads to scheduling unit of $180$kHz. Here, we use $540$kHz as the allocated bandwidth for URLLC traffic as an example.

\begin{figure}
  \centering
  \includegraphics[width=0.45\textwidth]{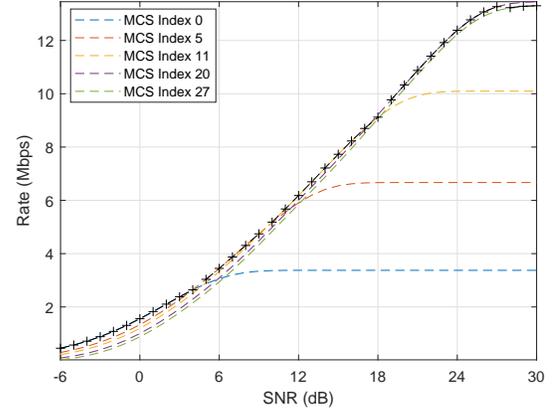}\\
  \caption{Transmission Rate  ($k=256$bits,  $\epsilon=10^{-3}$, $W=540$kHz)}\label{fig:rate103}
\end{figure}

Fig. \ref{fig:rate103} plots the transmission rate when $k=256$bits and $\epsilon=10^{-3}$ under different MCS schemes. It is obvious that the transmission rate will be significantly improved when SNR increases or more efficient MCS scheme is applied.
In addition, the curve with maker "$+$" is composed of those transmission rates after considering adaptive MCS selection with SNR thresholds given in Table. \ref{tbl:SNRthresholds}. It can be seen that optimal rates can be achieved after applying adaptive MCS selection. 

\begin{figure}
  \centering
  \includegraphics[width=0.45\textwidth]{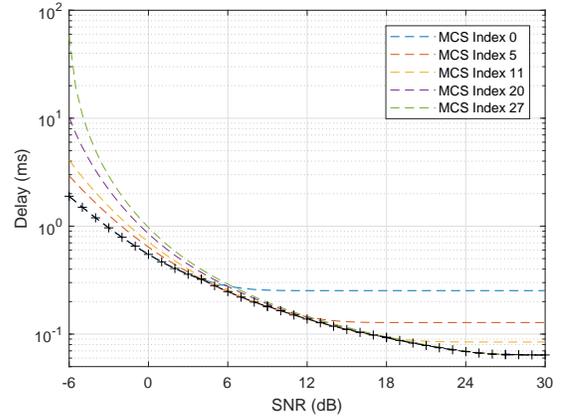}\\
  \caption{Maximum Delay  ($k=256$bits,  $\epsilon=10^{-3}$, $W=540$kHz)}\label{fig:delay103}
\end{figure}

For the considered periodical URLLC traffic, the maximum delay can be calculated out by Eq.(\ref{eq:delayforperi}) based on the transmission rate given in Fig. \ref{fig:delay103}. 

It can be noticed in Fig. \ref{fig:delay103} that, the maximum delay is more than $1$ms when SNR is below around $-3$dB.
If the periodical URLLC traffic can tolerate a maximum delay of $1$ms and error probability of $10^{-3}$, it means that more bandwidth should be allocated when SNR is below $-3$dB, while less bandwidth can be allocated  when SNR is above $-3$dB in order to guarantee its QoS requirement and at the same time to consume as less system resource as possible.

\begin{figure}
  \centering
  \includegraphics[width=0.5\textwidth]{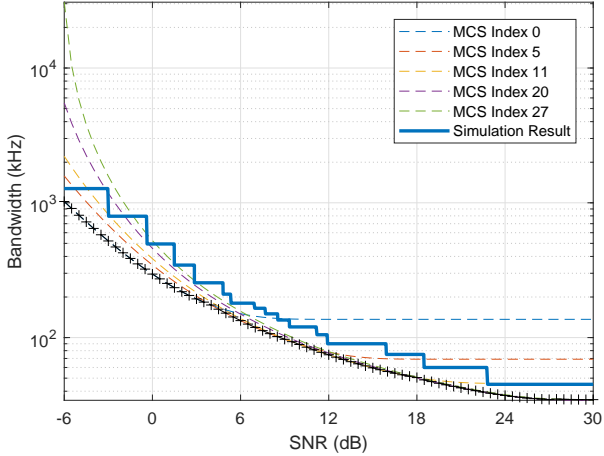}\\
  \caption{Capacity Region  ($k=256$bits,  $d_0=1$ms, $\epsilon_0=10^{-3}$)}\label{fig:bandwidth103}
\end{figure}

The minimum bandwidth under certain QoS requirement $(d_0,\epsilon_0)$ and certain SNR gives the lower bound of admission region as defined in Section III. Fig. \ref{fig:bandwidth103} shows the minimum required bandwidth under QoS constraint of $(d_0=1$ms, $\epsilon_0=10^{-3}$). The region above the  curve with marker $+$ forms the admission region with adaptive MCS selection. 

Simulation is also made to validate the theoretical analysis in this work. A simulation platform is established, where a typical deployment scenario is considered, i.e.,  an area of $19$ cells (further divided into $57$ sectors) with radius of $500$ meters. Users with periodical URLLC traffic are uniformed deployed within the considered area, and user density is $10$ users per sector. All MCS schemes defined in  \cite{TS38214} are used in the simulation, where the best suitable MCS scheme will be selected in order to guarantee the QoS constraint of $(d_0=1$ms, $\epsilon_0=10^{-3}$). The solid line in Fig. \ref{fig:bandwidth103} is formed of the simulation results. Each step indicates one MCS scheme. It can be seen that the simulation results are slightly higher than the theoretical results, because the theoretical results are the lowest bound under ideal scenario.


\section{Conclusion}
In this work, we have established a theoretical analysis model by considering all necessary factors during the transmission of a URLLC message, including channel quality indicated by SNR, finite coding length, modulation scheme, delay and reliability  requirement as well as information length. Network calculus is then applied to derive the maximum delay bound and minimum spectrum admission region. Configurations defined in 5G NR network are used to obtain numerical results, where adaptive MCS selection thresholds and admission region under certain delay and reliability constraint are presented and discussed. In addition, the theoretical analysis is validated by comparing with simulation results yielded from system level simulation platform.

%
%

%




\ifCLASSOPTIONcaptionsoff
  \newpage
\fi



\end{document}